\documentclass[nobibnotes,preprintnumbers,aps,prl,superscriptaddress,showpacs,twocolumn,twoside,sort&compress]{revtex4}				 %
\usepackage{graphicx}																												 %
\usepackage{titlesec}	
\usepackage{color}																										     %
\usepackage{fancyheadings}																											 %
\begin{document}																													 %
\preprint{{Vol.XXX (201X) ~~~~~~~~~~~~~~~~~~~~~~~~~~~~~~~~~~~~~~~~~~~~~~~~~~~~ {\it CSMAG`16}										  ~~~~~~~~~~~~~~~~~~~~~~~~~~~~~~~~~~~~~~~~~~~~~~~~~~~~~~~~~~~~ No.X~~~~}}																 %
\vspace*{-0.3cm}																													 %
\preprint{\rule{\textwidth}{0.5pt}}																											 \vspace*{0.3cm}																														 %

\title{Rise and Fall of Reentrant Phase Transitions in a Coupled Spin-Electron Model \\ on a Doubly Decorated Honeycomb Lattice}

\author{H. \v{C}en\v{c}arikov\'a}
\thanks{corresponding author; e-mail:hcencar@saske.sk}
\affiliation{Institute of Experimental Physics SAS, Watsonova 47, 040 01 Ko\v{s}ice, Slovakia}
\author{J. Stre\v{c}ka}
\affiliation{Institute of Physics, Faculty of Science, P. J. \v{S}af\'arik University, Park Angelinum 9, 040 01 Ko\v{s}ice, Slovakia}

\begin{abstract}
Phase diagrams and spontaneous magnetization are rigorously calculated for a coupled spin-electron model on a doubly decorated honeycomb lattice, which accounts for a quantum-mechanical hopping of the mobile electrons on decorating sites, the nearest-neighbor Ising coupling between mobile electrons and localized spins, as well as, the further-neighbor Ising coupling between the localized spins placed on nodal sites. The spontaneously ordered ferromagnetic phase, spontaneously ordered antiferromagnetic phase and disordered paramagnetic phase emerge in a phase diagram depending on an electron filling of the decorating sites, a relative size of the hopping term and both considered coupling constants. It is evidenced that a nature and size of the further-neighbor Ising coupling between the localized spins basically influences rise and fall of reentrant transitions close to a phase boundary between the paramagnetic phase and both spontaneously ordered phases. 
\end{abstract}


\maketitle

%
\section{Introduction}
A reentrant transition between the same phases is a fascinating cooperative phenomenon, which has been experimentally observed in a variety of physical systems including spin glasses, liquid mixtures, superconductors or intermetallic compounds \cite{nara94}. The spectrum of magnetic reentrant transitions is very rich and theoretical models for a description of this remarkable phenomenon are therefore highly desirable. Recently, the reentrant phase transitions have been thoroughly investigated in a coupled spin-electron model on doubly decorated planar lattices \cite{dori14,cenci1}, whereas the follow-up investigations demonstrated a crucial role of the further-neighbor interaction between the localized spins on a presence or absence of the reentrance in the coupled spin-electron model on a doubly decorated square lattice \cite{cenci2}. The main goal of the present work is to explore rise and fall of the reentrant phase transitions in the analogous spin-electron model on a doubly decorated honeycomb lattice, which will enable us to clarify an influence of the underlying lattice geometry.    
\begin{figure}[h!]
\begin{center}
\includegraphics[width=6.5cm,height=4cm]{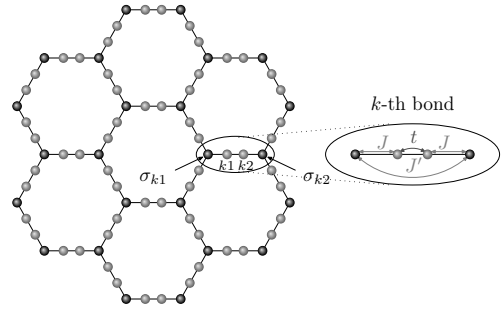}
\vspace*{-0.2cm}
\caption{\small A schematic representation of the coupled spin-electron model on a doubly decorated honeycomb lattice. Black balls represent nodal sites occupied by localized spins, while gray balls denote decorating sites available to mobile electrons. The ellipse illustrates the $k$-th bond.}
\label{fig1}
\end{center}
\end{figure}
\section{Model and method}
Let us consider the coupled spin-electron model on the doubly decorated honeycomb lattice illustrated in Fig.~\ref{fig1}. The investigated model involves one localized spin at each nodal site (black ball) and variable number of mobile electrons on decorating sites (grey balls). The total Hamiltonian can be defined as a sum of the bond Hamiltonians $\hat{\cal H}=\sum_{k=1}^{Nq/2}\hat{\cal H}_k$, which are defined as follows
\begin{eqnarray}
\hat{\cal H}_k= \!\!\!&-&\!\!\! t(\hat{c}^\dagger_{k1,\uparrow}\hat{c}_{k2,\uparrow}+\hat{c}^\dagger_{k1,\downarrow}\hat{c}_{k2,\downarrow}+
\hat{c}^\dagger_{k2,\uparrow}\hat{c}_{k1,\uparrow}+\hat{c}^\dagger_{k2,\downarrow}\hat{c}_{k1,\downarrow})
\nonumber
\\
\!\!\!&-&\!\!\! J\hat\sigma^z_{k1}(\hat{n}_{k1,\uparrow}-\hat{n}_{k1,\downarrow})-
J\hat\sigma^z_{k2}(\hat{n}_{k2,\uparrow}-\hat{n}_{k2,\downarrow})
\nonumber\\
\!\!\!&-&\!\!\! J'\hat\sigma^z_{k1}\hat\sigma^z_{k2}.
\label{eq1}
\end{eqnarray}
In above, the symbols $\hat{c}^\dagger_{k\alpha,\gamma}$ and $\hat{c}_{k\alpha,\gamma}$ ($\alpha$=1,2, $\gamma$=$\uparrow,\downarrow$) denote creation and annihilation fermionic operators for the mobile electron with the spin $\gamma$, whereas $\hat{n}_{k\alpha,\gamma}=\hat{c}^\dagger_{k\alpha,\gamma}\hat{c}_{k\alpha,\gamma}$ is the respective number operator. The spin operator $\hat\sigma^z_{i}$ relevant to the $i$-th nodal site denotes $z$-component of the spin-1/2 operator with the corresponding eigenvalues $\sigma^z_{i}=\pm1$. The first term $t$ accounts for the quantum-mechanical hopping of the mobile electrons delocalized over a couple of the decorating sites, the second term $J$ represent the Ising coupling between the mobile electrons and their nearest localized spins, and the last term $J'$ corresponds to the Ising coupling between the nearest-neighbor localized spins. Finally, $N$ is the total number of the nodal sites and $q$ is their coordination number ($q=3$ for the present case).

The coupled spin-electron model on doubly decorated planar lattices defined through the Hamiltonian (\ref{eq1}) has been exactly solved in our preceding work \cite{cenci2}, to which readers interested in calculation details are referred to. Hereafter, our attention will be exclusively focused on a discussion of the most interesting results obtained for phase diagrams and spontaneous magnetizations of the coupled spin-electron model on the doubly decorated honeycomb lattice.  

\section{Results and discussion}
Before presenting the most interesting results we would like to stress that one may consider without loss of generality the ferromagnetic coupling $J>0$ between the localized spins and mobile electrons, because the critical temperature and other thermodynamic quantities (except the sign of order parameters) remain unchanged under the transformation $J\to-J$. In addition, the electron density per one decorating dimer can be also restricted just up to a half filling $\rho \leq 2$ due to a validity of the particle-hole symmetry. For simplicity, we will use the magnitude of the nearest-neighbor Ising interaction $J>0$ as the energy unit when normalizing all other parameters with respect to this coupling constant.

\begin{figure}[h!]
\vspace*{-0.5cm}
\includegraphics[width=0.9\columnwidth]{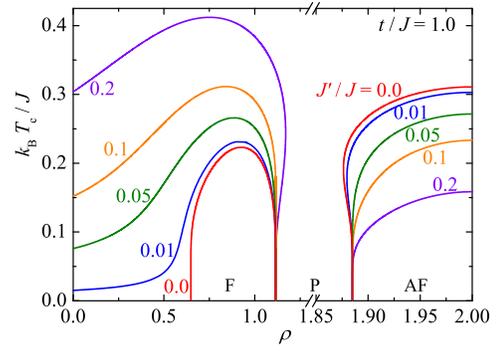}
\vspace*{-0.7cm}
\caption{The phase diagram in the critical temperature vs. electron density plane for a relative size of the hopping term $t/J = 1.0$ 
and several values of the F interaction $J'/J>0$ between the localized spins.}
\label{fig2}
\end{figure}

Fig.~\ref{fig2} sheds light on an influence of the ferromagnetic interaction $J'>0$ between the localized spins on a finite-temperature phase diagram, which involves the spontaneously ordered ferromagnetic (F) phase, the spontaneously ordered antiferromagnetic (AF) phase and the disordered paramagnetic (P) phase. If the further-neighbor coupling between the localized spins is absent $J'=0$, then, the ground state is formed by the F phase for the electron densities around a quarter filling $\rho \in (0.648, 1.115)$, the AF phase for the electron densities around a half filling $\rho \in (1.885, 2)$ and the P phase for the rest of electron densities. It is quite obvious from Fig.~\ref{fig2} that the F ground state extends down to zero electron density ($\rho=0$) for any non-zero ferromagnetic coupling $J'>0$, whereas the respective critical temperature generally increases upon strengthening of $J'$. Another interesting observation is that the double reentrant transitions between the F and P phase can be detected close to but slightly above the electron density $\rho \gtrsim 1.115$ whenever the interaction $J'$ between the localized spins is sufficiently strong. On the contrary, the critical temperature of the $AF$ phase is suppressed upon increasing of the ferromagnetic interaction $J'>0$, which additionally causes a gradual breakdown of the reentrance between the AF and P phase.

\begin{figure}[h!]
\vspace*{-0.5cm}
\hspace*{-3.6cm}
\includegraphics[width=0.85\columnwidth]{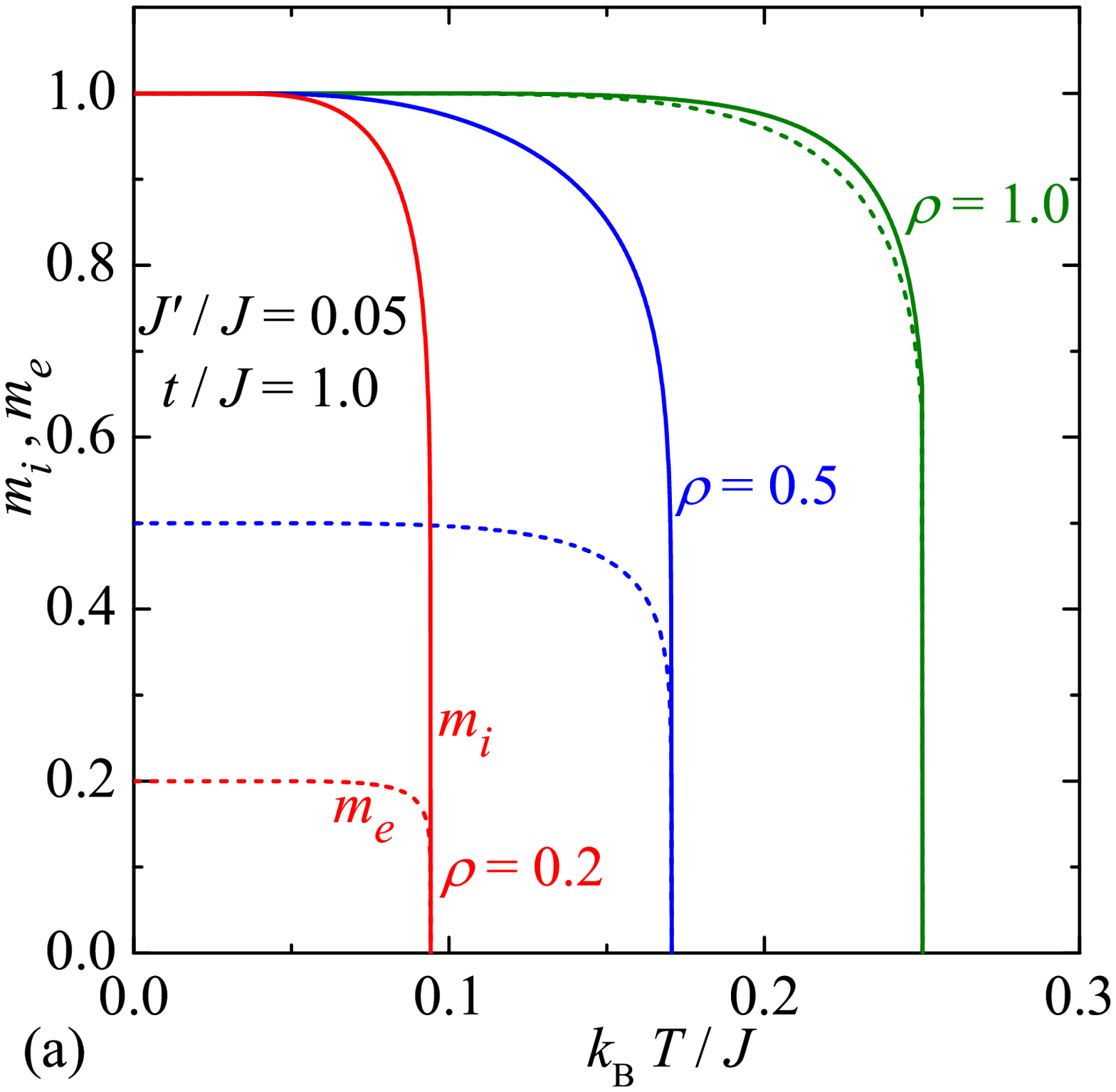}
\hspace*{-3.2cm}
\includegraphics[width=0.85\columnwidth]{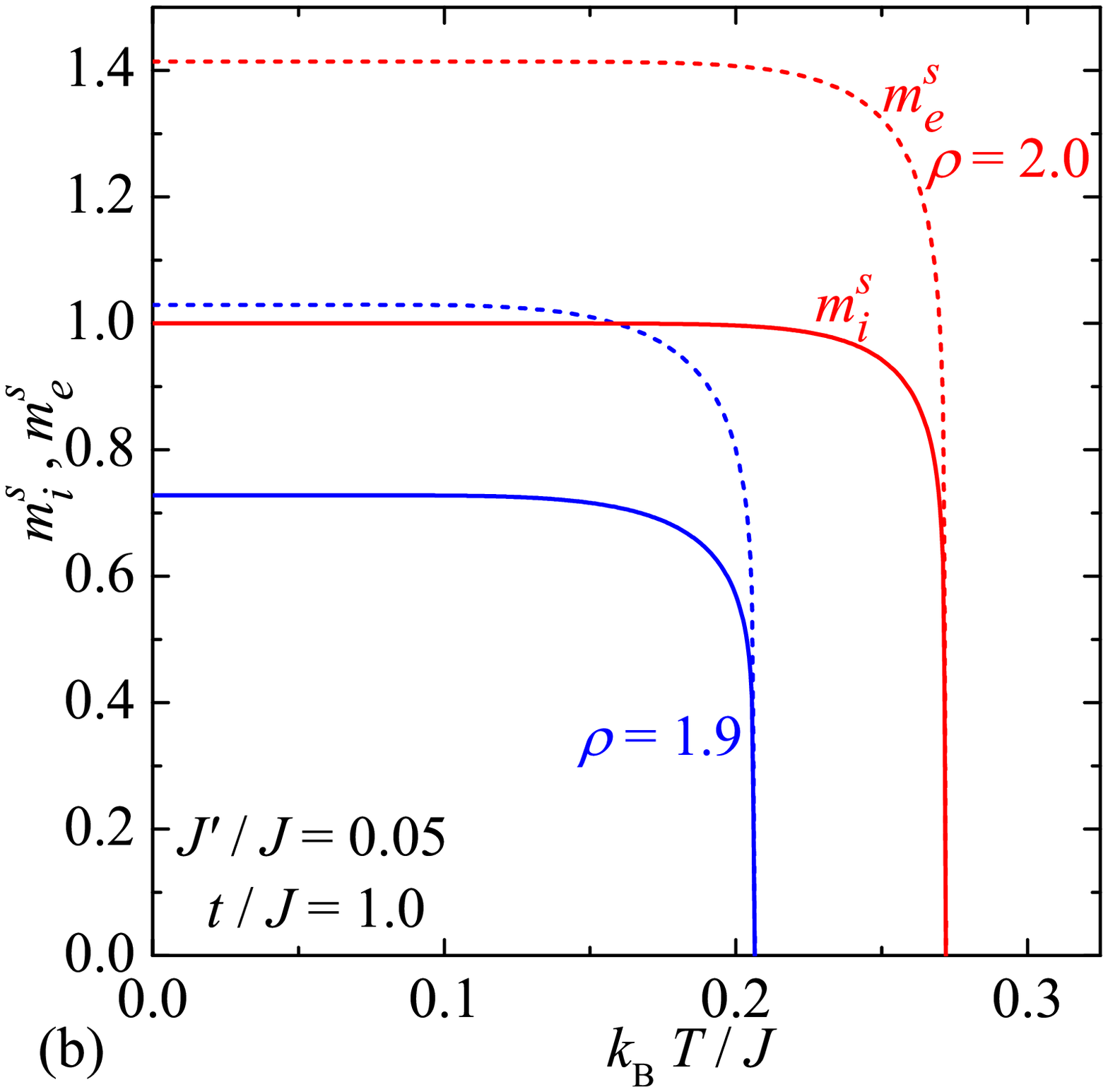}
\hspace*{-5.4cm}
\vspace*{-0.7cm}
\caption{(a) Thermal variations of spontaneous uniform magnetizations of localized spins (solid lines) and mobile electrons (dashed lines) for $t/J=1.0$, $J'/J=0.05$ and three different values of the electron density; (b) Thermal variations of spontaneous staggered magnetizations of localized spins (solid lines) and mobile electrons (dashed lines) for $t/J=1.0$, $J'/J=0.05$ and two different values of the electron density.}
\label{fig3}
\end{figure}

To gain a deeper insight into a nature of the spontaneous ordering, we have depicted in Fig.~\ref{fig3} a few typical thermal variations of the uniform (Fig.~\ref{fig3}(a)) and staggered (Fig.~\ref{fig3}(b)) magnetizations verifying the spontaneous F and AF long-range order, respectively. It is quite obvious from Fig.~\ref{fig3}(a) that the zero-temperature asymptotic limit of the spontaneous uniform magnetization of the localized Ising spins $m_i$ tends within the F phase towards its saturation value quite similarly as does the spontaneous uniform magnetization of the mobile electrons $m_e$, whose saturation value is however suppressed upon decreasing of the electron density. On the other hand, it is clear from Fig.~\ref{fig3}(b) that the spontaneous staggered magnetization of the mobile electrons $m_e^s$ never reaches within the AF phase its saturation value due to a mutual interplay between the quantum reduction of magnetization and annealed bond disorder, while the spontaneous staggered magnetization of the localized spins $m_i^s$ reaches its saturation value just at a half filling $\rho=2$ owing to an absence of the annealed bond disorder. 

\begin{figure}[h!]
\vspace*{-0.5cm}
\includegraphics[width=0.9\columnwidth]{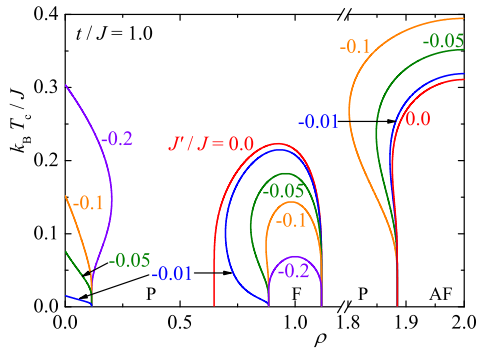}
\vspace*{-0.7cm}
\caption{The phase diagram in the critical temperature vs. electron density plane for a relative size of the hopping term $t/J = 1.0$ 
and several values of the AF interaction $J'/J<0$ between the localized spins.}
\label{fig4}
\end{figure}

\begin{figure}[h!]
\vspace*{-0.5cm}
\hspace*{-3.6cm}
\includegraphics[width=0.85\columnwidth]{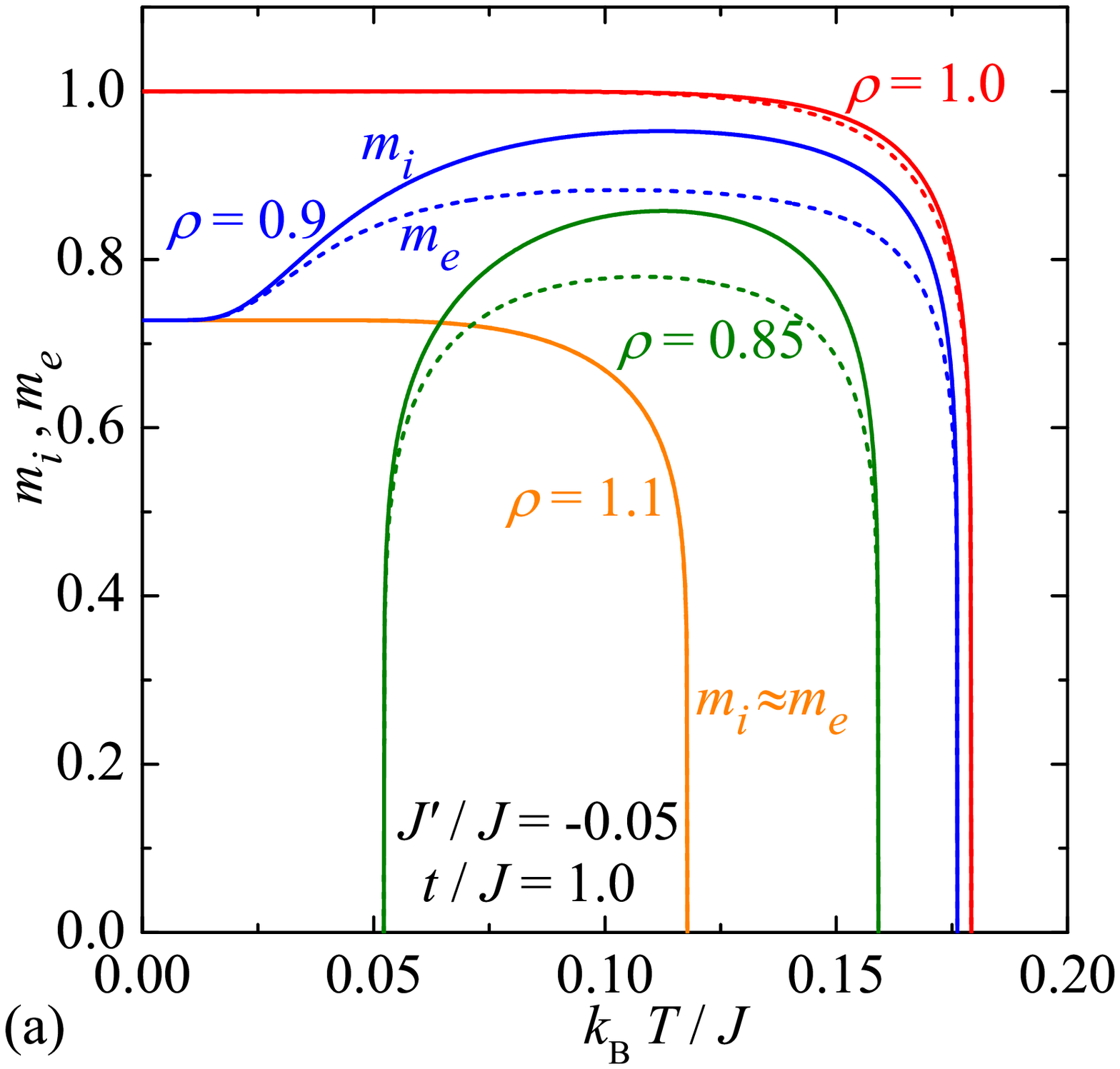}
\hspace*{-3.2cm}
\includegraphics[width=0.85\columnwidth]{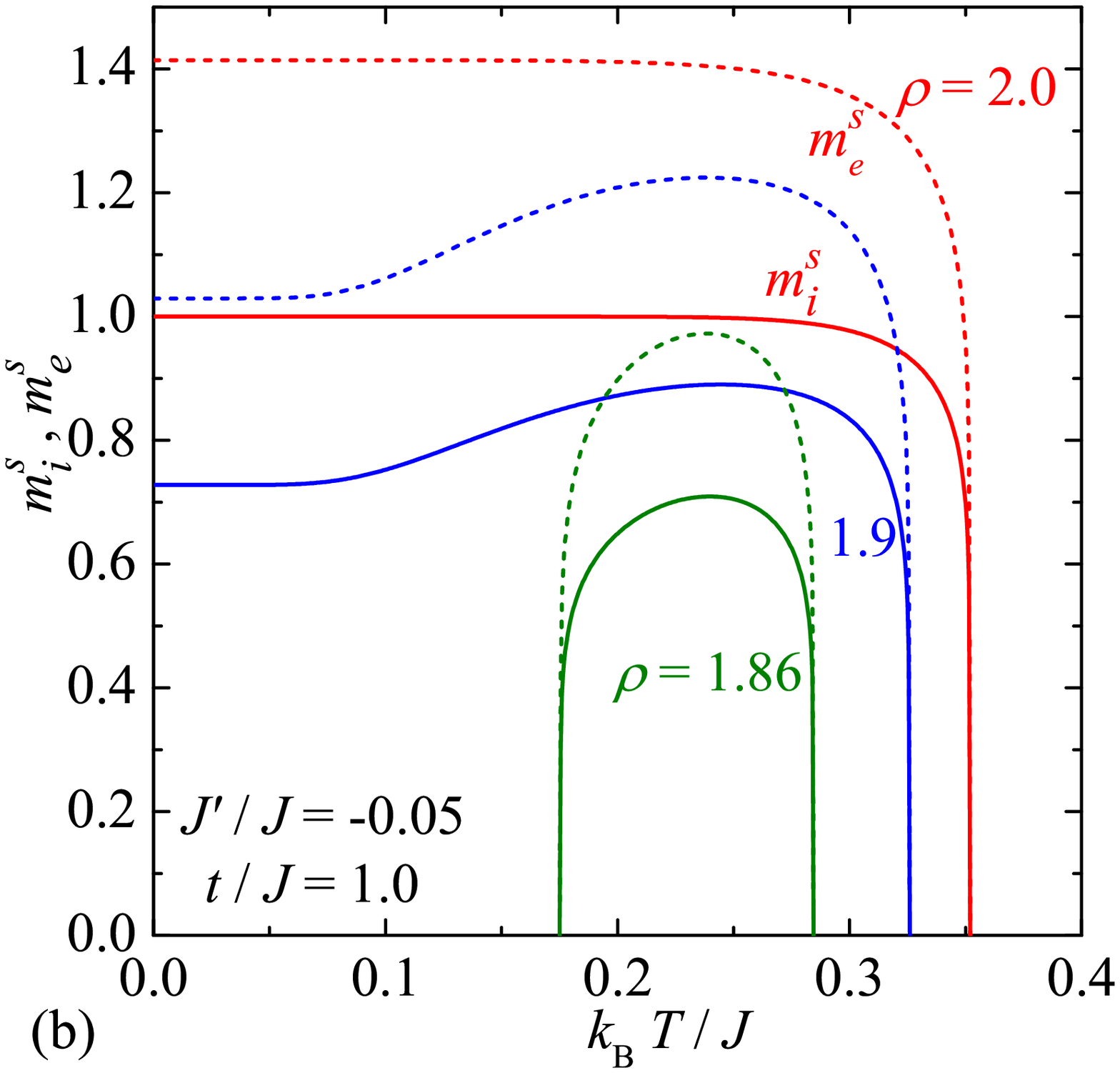}
\hspace*{-5.4cm}
\vspace*{-0.7cm}
\caption{(a) Thermal variations of spontaneous uniform magnetizations of localized spins (solid lines) and mobile electrons (dashed lines) for $t/J=1.0$, $J'/J=-0.05$ and several values of the electron density; (b) Thermal variations of spontaneous staggered magnetizations of localized spins (solid lines) and mobile electrons (dashed lines) for $t/J=1.0$, $J'/J=-0.05$ and three different values of the electron density.}
\label{fig5}
\end{figure}

Next, let us turn our attention to an effect of the AF interaction $J'<0$ between the localized spins upon the finite-temperature phase diagram. It could be easily understood from Fig.~\ref{fig4} that the completely reverse trends are observed: the spontaneous F order is suppressed and the spontaneous AF order is reinforced upon strengthening of the AF interaction $|J'|$. In fact, the ground state shows the spontaneous F order just in a very limited range of the electron densities $\rho \in (0.885, 1.115)$, whereas the pronounced double reentrant transitions between the F and P phase can be detected at the electron densities $\rho \lesssim 0.885$ when assuming a sufficiently weak AF interaction $|J'|$. By contrast, the spontaneous AF order does occur at zero temperature either for small enough $\rho \in (0,0.115)$ or large enough $\rho \in (1.885,2)$ electron densities, whereas the respective phase boundaries are broadened upon strengthening of $|J'|$. Owing to this fact, the double reentrant  transitions between the AF and P phase can be observed either for $\rho \gtrsim 0.115$ or $\rho \lesssim 1.885$. 

To support this statement, we have displayed in Fig.~\ref{fig5} temperature dependences of the spontaneous uniform (Fig.~\ref{fig5}(a)) and staggered (Fig.~\ref{fig5}(b)) magnetizations within the F and AF phase for the AF interaction $J'<0$ between the localized spins. Unlike the previous case, the spontaneous uniform magnetizations of the localized spins $m_i$ and mobile electrons $m_e$ tend to the respective saturation values only at a quarter filling $\rho=1$, while they exhibit the striking thermally-induced increase out from the quarter-filling case. Hence, it follows that the annealed bond disorder in conjunction with the AF further-neighbor interaction $J'<0$ destabilize the spontaneous F long-range order, which may be however recovered by thermal fluctuations as evidenced by loop thermal dependences of the spontaneous uniform magnetizations showing the double reentrant phase transitions (see the curve $\rho = 0.85$ in Fig.~\ref{fig5}(a)). The remarkable temperature-driven increase of the spontaneous staggered magnetization of the localized spins $m_i^s$ and mobile electrons $m_e^s$ can be also found in the AF phase whenever the electron density deviates from a half filling $\rho=2$ (Fig.~\ref{fig5}(b)). In agreement with previous argumentation, the staggered magnetizations of the localized spins and mobile electrons display an outstanding loop dependence with the double reentrant phase transitions between the AF and P phase for the electron densities $\rho \lesssim 1.885$ (see the curve $\rho = 1.86$ in Fig.~\ref{fig5}(b)).


In conclusion, we have examined the phase diagrams and order parameters of the coupled spin-electron model on the doubly decorated honeycomb lattice, which exhibits the spontaneously ordered F phase, the spontaneously ordered AF phase and the disordered P phase. Interestingly, the investigated model displays under certain conditions the double reentrant transitions between the spontaneously ordered phase (either F or AF) and the disordered P phase. In addition, it has been convincingly evidenced that the double reentrance crucially depends on a character and relative size of the coupling constant $J'$ between the localized spins. It actually turns out that the F (AF) further-neighbor interaction $J'>0$ ($J'<0$) preferentially gives rise to the reentrant transitions of the F (AF) phase, while it conversely inhibits the reentrant transitions of the AF (F) phase. 

\section{Acknowledgement}
This work was financially supported under the grant Nos. VEGA 1/0043/16 and APVV-0097-12.

\end{document}